\documentclass[a4paper,english,reprint, aps,onecolumn]{revtex4-2}
\usepackage[T1]{fontenc}
\usepackage[latin9]{inputenc}
\usepackage{color}
\usepackage{array}
\usepackage{multirow}
\usepackage{amsmath}
\usepackage{amssymb}
\usepackage{graphicx}

\makeatletter

\newcommand{\lyxmathsym}[1]{\ifmmode\begingroup\def\b@ld{bold}
  \text{\ifx\math@version\b@ld\bfseries\fi#1}\endgroup\else#1\fi}

\providecommand{\tabularnewline}{\\}

\makeatother

\usepackage{babel}
\begin{document}
 
\title{Solar Flare 1/f Fluctuations from Amplitude Modulated Five Minute
Oscillation}
\author{Masahiro Morikawa}
\email{hiro@phys.ocha.ac.jp}

\affiliation{Department of Physics, Ochanomizu University ~\\
2-1-1 Otsuka, Bunkyo, Tokyo 112-8610, Japan }
\author{Akika Nakamichi}
\email{nakamichi@cc.kyoto-su.ac.jp}

\affiliation{General Education, Kyoto Sangyo University ~\\
Motoyama Kamigamo Kita-ku, Kyoto 603-8555 Japan }
\date{\today}
\begin{abstract}
We first study the solar flare time sequence based on the GOES16 data.
We find that the power spectrum density of the low-energy ($E\leq E_{mean}$)
flare shows 1/f fluctuations, but the high-energy ($E>E_{mean}$)
flare shows a flat spectrum. Further, we found that the flare timing
time-sequence shows 1/f fluctuations clearer. These facts indicate
that the solar flare 1/f fluctuations are associated with low-energy
phenomena. We investigate the origin of this 1/f fluctuations based
on our recent proposal: 1/f fluctuations arise from amplitude modulation
and demodulation. We speculate that this amplitude modulation is
encoded by the resonance with the Solar Five-minute Oscillation (SFO)
and demodulated by magnetic reconnection.We partially demonstrate
this scenario by analyzing the SFO eigenmodes resolving the frequency
degeneracy in the azimuthal order number $m$ by solar rotation and
resonance. Since 1/f fluctuation is robust, we speculate that the
solar flare 1/f fluctuations may be inherited by the various phenomena
around the Sun, such as the sunspot numbers and the cosmic rays. Finally,
we compare the solar flares and the earthquakes, both showing 1/f
fluctuations. Interestingly, the same analysis for solar flares is
possible for earthquakes if we read SFO as Earth's Free Oscillation,
and magnetic reconnection as fault rupture. Furthermore, we point
out the possibility that the same analysis also applies to the activity
of the black hole/disk system if we read SFO as the Quasi-Periodic
Oscillation of a black hole. 

\end{abstract}

\maketitle

\section{Introduction}

A Solar flare is a sudden energy eruption in the solar atomosphere\citep{Benz2008}.
It is triggered by magnetic reconnection, and the enormous magnetic
energy $10^{17}-10^{26}$ J is converted into plasma particle acceleration,
heating, and light emission. Solar flares are quite complex phenomena,
and the statistical approach is effective, as in the case of earthquakes,
being a sudden energy eruptions in the Earth's crust. 

It is well known that solar flares and earthquakes are similar to
each other, and they show similar statistical properties. In particular,
scaling relations, such as the Gutenberg-Richter law\citep{Gutenberg1944}
and the Omori law\citep{Omori1894}, are universal laws for both solar
flares and earthquakes \citep{Arcangelis2006,Najafi2020}. 

Here in this paper, we concentrate on the solar flare and would like
to add one more universal scaling law in the ultra-low frequency region
of the power spectrum density (PSD) for the long time-sequnce of the
solar flares. It turns out that the solar flare time sequence shows
the power law almost inversely proportional to the frequency in PSD.
This is often called 1/f fluctuation or pink noise and appears in
most fields in nature and human activities\citep{Raychaudhuri2002,Milott2002}.
However, the origin of this fluctuation was not clarified despite
tremendous studies in the past century. 

We recently proposed that the general origin of pink noise is amplitude
modulation (AM), or the beat of many waves with accumulating frequencies\citep{Morikawa2023}.
In particular, this frequency accumulation is possible in resonance
where many eigenfrequencies are systematically concentrated in a narrow
domain. 

Applying this method in \citep{Nakamichi2023}, we studied pink noise
in the seismic activities. The seismic-energy time sequence shows
an apparent pink noise in its PSD in more than three digits if giant
earthquakes are excluded. Therefore, seismic pink noise is considered
to be associates with low-energy phenomena. In this case, perpetually
exciting Earth Free Oscillation (EFO) in the lithosphere is resonating
to yield AM or wave beats. Relatively low energy EFO will sufficiently
trigger the fault ruption and cause earthquakes. 

In this paper, applying this proposal to solar flares, we would like
to verify our proposal and try to figure out the statistical properties
of the complex systems in general. When we naively use all the energy
time series data, we obtain almost flat PSD at low-frequency regions
and obtain no clear pink noise. However, if we restrict the solar
flare events with their energy below the mean, we obtain clear pink
noise. Therefore, we speculate that the pink noise in solar flares
is associated with low-energy phenomena. Interestingly, this situation
is the same as the seismic activity, as explained above. 

The resonation in the solar case would be characterized by the solar
five-minute oscillations (SFO), which is perpetually excited by the
turbulence in the solar convective region\citep{Leighton1962,Evans1962}.
The eigenfrequencies are precisely measured and calculated by many
studies assuming appropriate solar models. Using this observational
data, we construct the superposition of waves with these eigen-oscillations,
including the fine splitting structure by the solar rotation and the
resonance effects. Then, we can obtain pink noise from the square
of this data in PSD. Thus, we can partially demonstrate the AM theory
for pink noise in solar flare. 

Since pink noises often appear in any field of science, we explore
the neighboring phenomena to the solar flare. Then we find several
phenomena such as solar wind, sun spot numbers, and some traces on
Earth. These facts may indicate the robustness of pink noises. 

The construction of this paper is as follows. In the next section
2, we explore the GOES data of solar flares and analyze PSD. In section
3, we explain the AM proposal for the origin of pink noise from resonance.
In section 4, we superpose the eigenmodes of SFO and obtain pink noise.
In section 5, we study another statistical characterization of the
solar flare by the Weibell distribution and compare it with the 1/f
characterization. In section 6, we emphasize the robustness of pink
noise is the origin of the variety of pink noises. We point out that
the pink noise property is inherited by solar winds, sunspot events,
and some traces on Earth. In section 7, we conclude our work and briefly
describe the possible future research. 

\section{Solar flare fluctuations \label{II. Solar flare fluctuations}}

 Solar flares are eruptive energy-release events in the solar atmosphere.
In each event, enormous magnetic energy is transformed into plasma
particle acceleration, visible light, X-ray, etc. Our focus is specifically
directed towards the soft X-ray flux data acquired by the GOES16 satellite
during the timeframe spanning from February 2017 to September 2023,
encompassing a duration of 6.6 years.\citep{GOES2023}. 

\begin{figure}
\textcolor{red}{\includegraphics[width=9cm]{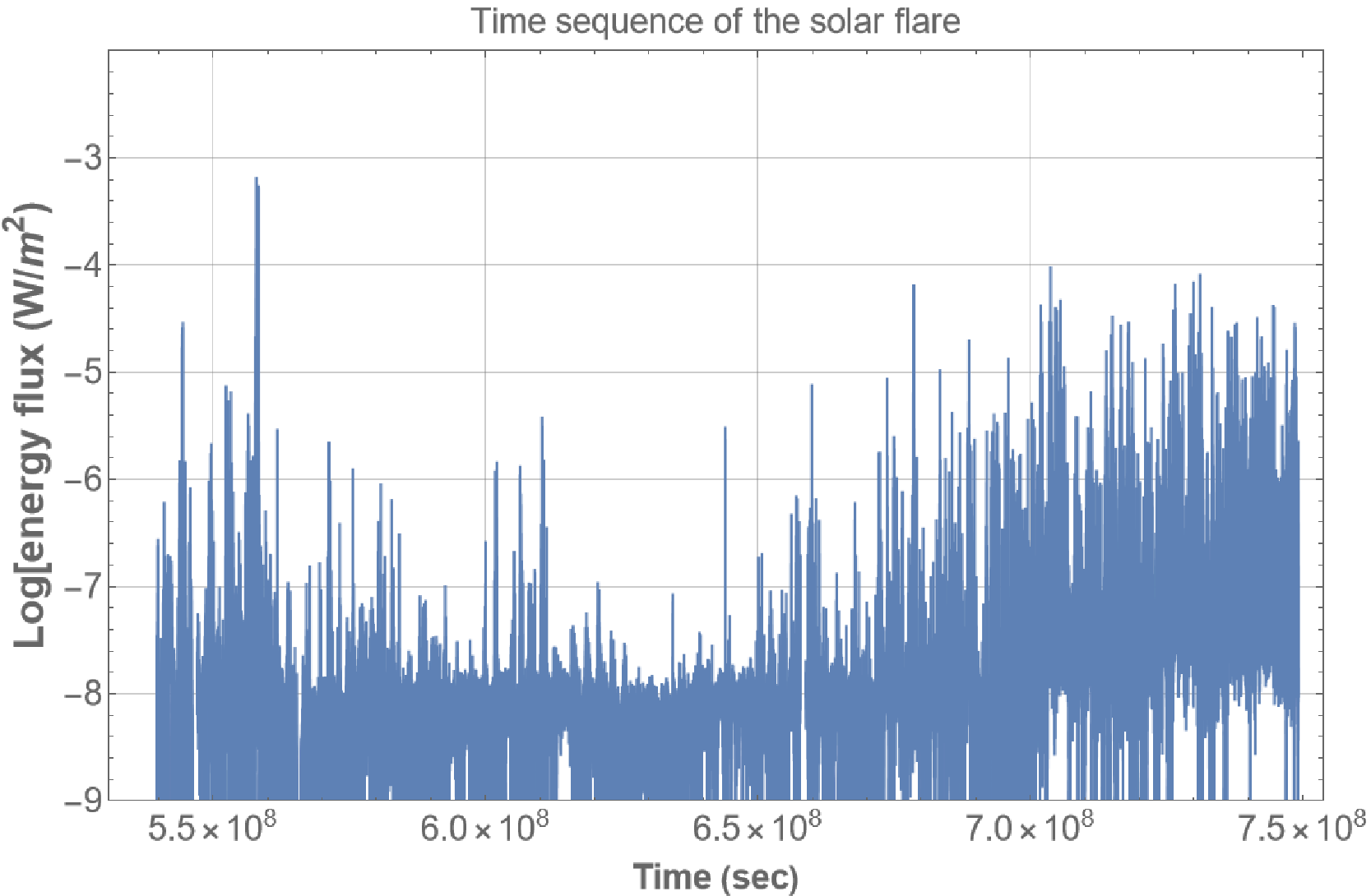}\includegraphics[width=9cm]{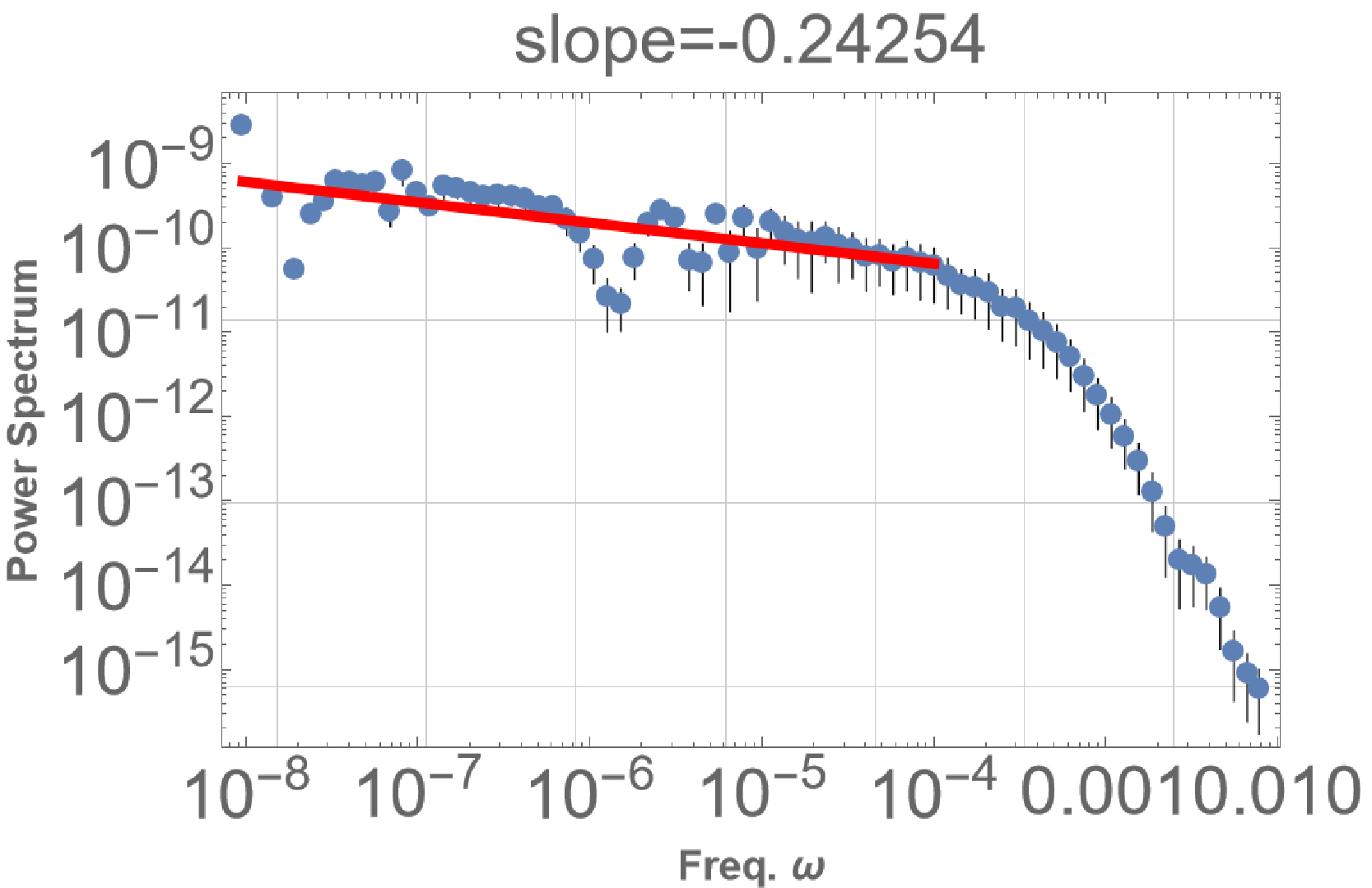}}\caption{left) Goes16 soft X-ray flux data (February 2017 to September 2023,
6.6 years)\citep{GOES2023}. The unit is $\mathrm{W/m^{2}}$ in the
logarismic scale. Although the data shows an apparent trend associated
with the eleven-year solar activity, we do not apply any artificial
operation for our analysis; extraction of the trend does not greatly
affect the result. \protect \\
right) Power Spectrum Density (PSD) of the energy-flux time-sequence
of all the GOES16 solar flare data 6.6 years. The time is measured
in seconds, and the frequency unit is Hz. Before the analysis, the
data are homogenized over time. This is the same for all PSD analyses
below. This PSD shows that the energy time sequence is random, as
shown by the almost flat red line that fits the data points in the
low-frequency domain. }
\label{fig1}
\end{figure}

We first naively use all the time sequence data of the soft X-ray
energy flux in the unit $\mathrm{W/m^{2}}$, Fig. \ref{fig1}left.
The corresponding PSD becomes almost flat and random in the low-frequency
regions, as in Fig. \ref{fig1} right. However, this is consistent
with the previous research\citep{Ueno1997}, in which the authors
of \citep{Ueno1997} partially extract 1/f fluctuations in the GOES6
data by superposing multiple PSDs. We try another approach to extract
the entire 1/f fluctuation by restricting energy flux. 

\begin{figure}
\includegraphics[width=9cm]{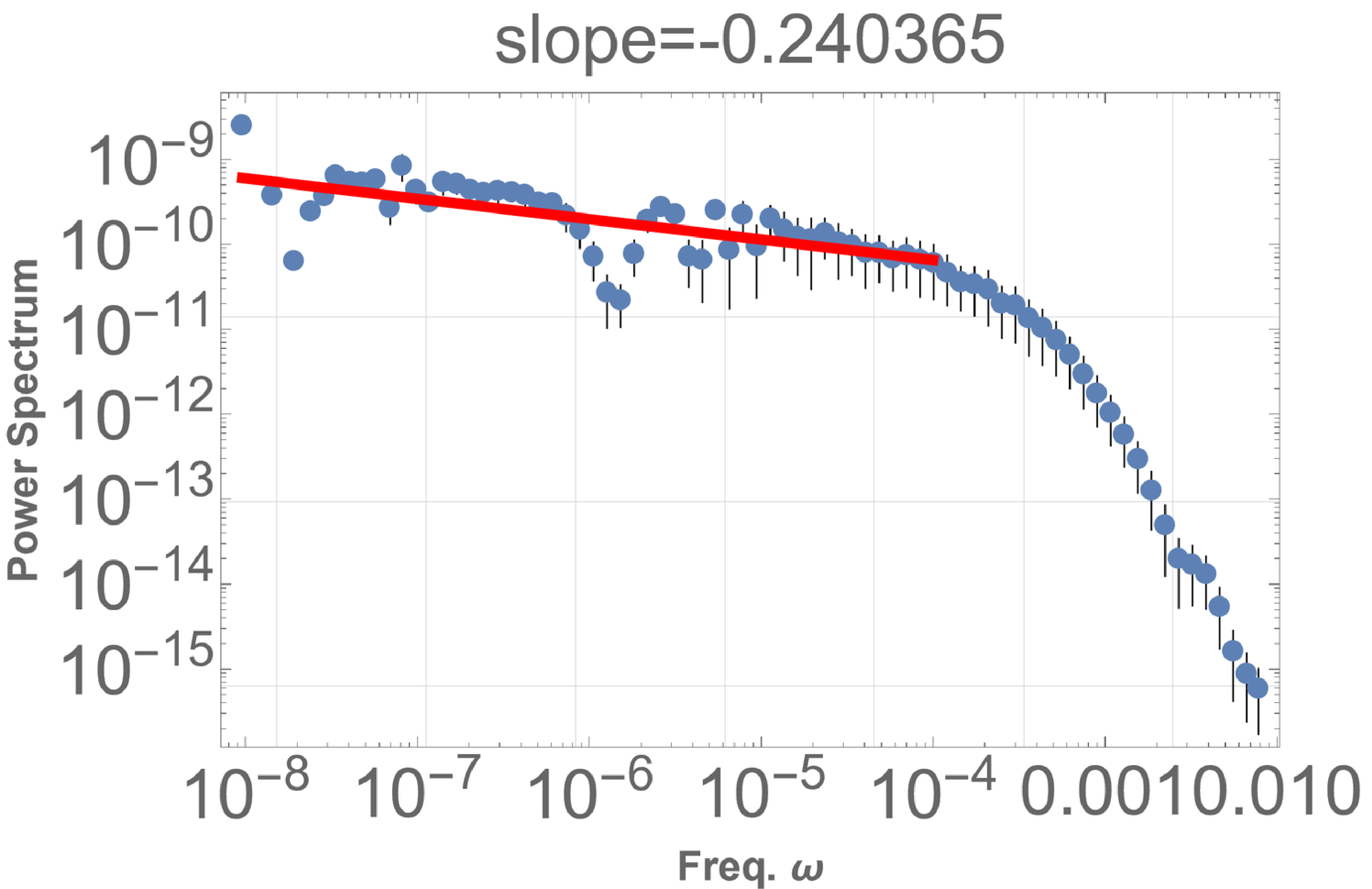}\includegraphics[width=9cm]{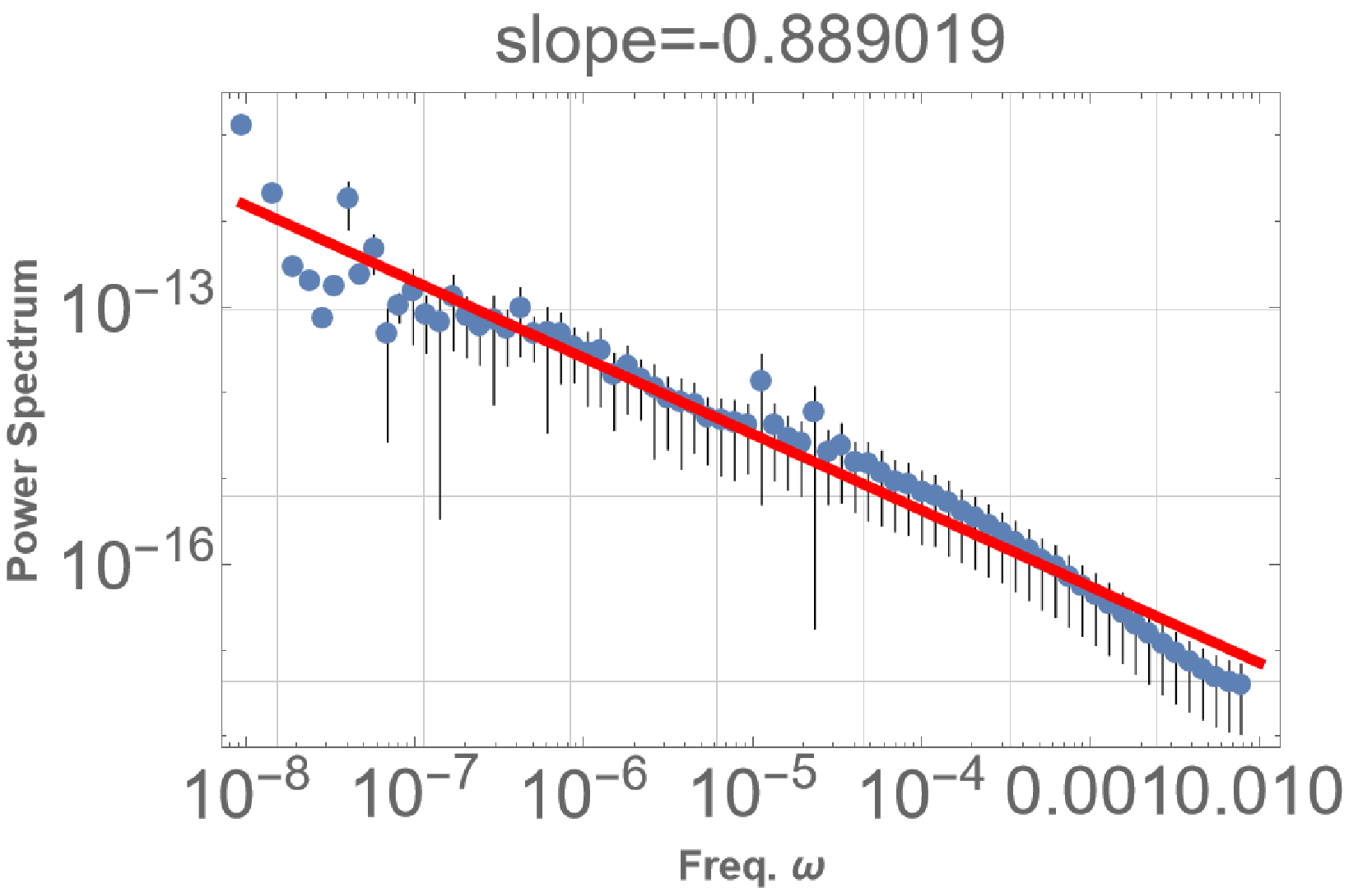}

\caption{left) PSD for the high-energy group, which include all the event with
the energy larger than the mean. The PSD does not show 1/f fluctuations,
similar to Fig.\ref{fig1} right.\protect \\
right) PSD for the low-energy group, which include all the event with
the energy smaller than the mean. The PSD shows clear pink noise with
an index of -0.89 in more than five digits.}
\label{fig2}
\end{figure}

Next, we do the same analysis for the two data sets: the high-energy
group, which include all the event with energy larger than the mean,
and the low-energy group, which include all the event with energy
smaller than the mean. PSDs are shown for these two data sets in Fig.
\ref{fig2}. It is apparent that the pink noise does appear in Fig.\ref{fig2}
right, the low-energy data. On the other hand, PSD for the high-energy
group does not show 1/f fluctuations as in Fig.\ref{fig2} left, similar
to Fig.\ref{fig1} righr; high-energy data destroys the 1/f fluctuations
in the solar flare. These facts indicate that the solar flare pink
noise is associated with low-energy phenomena. 

\begin{figure}
\includegraphics[width=9cm]{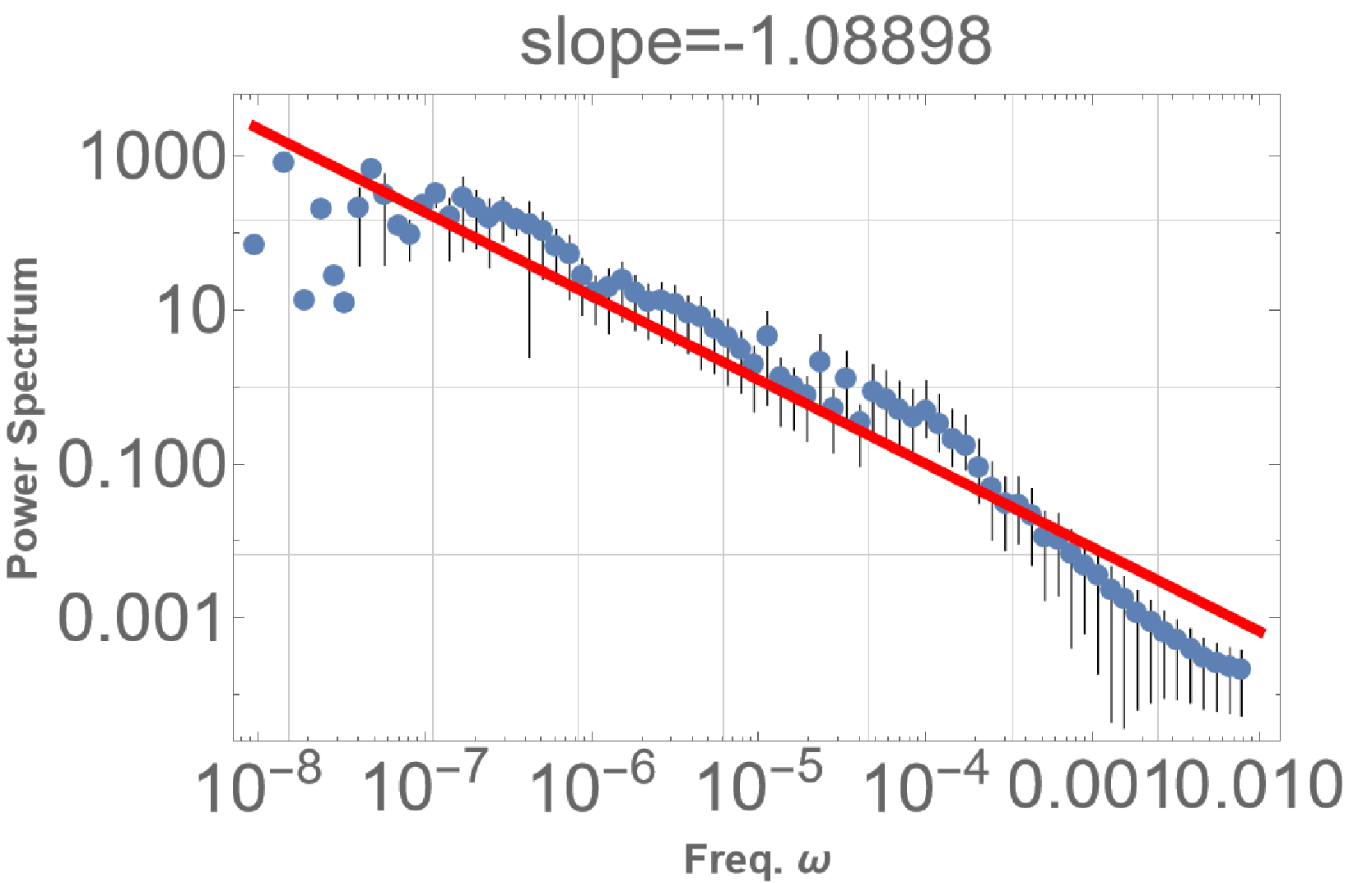}\includegraphics[width=9cm]{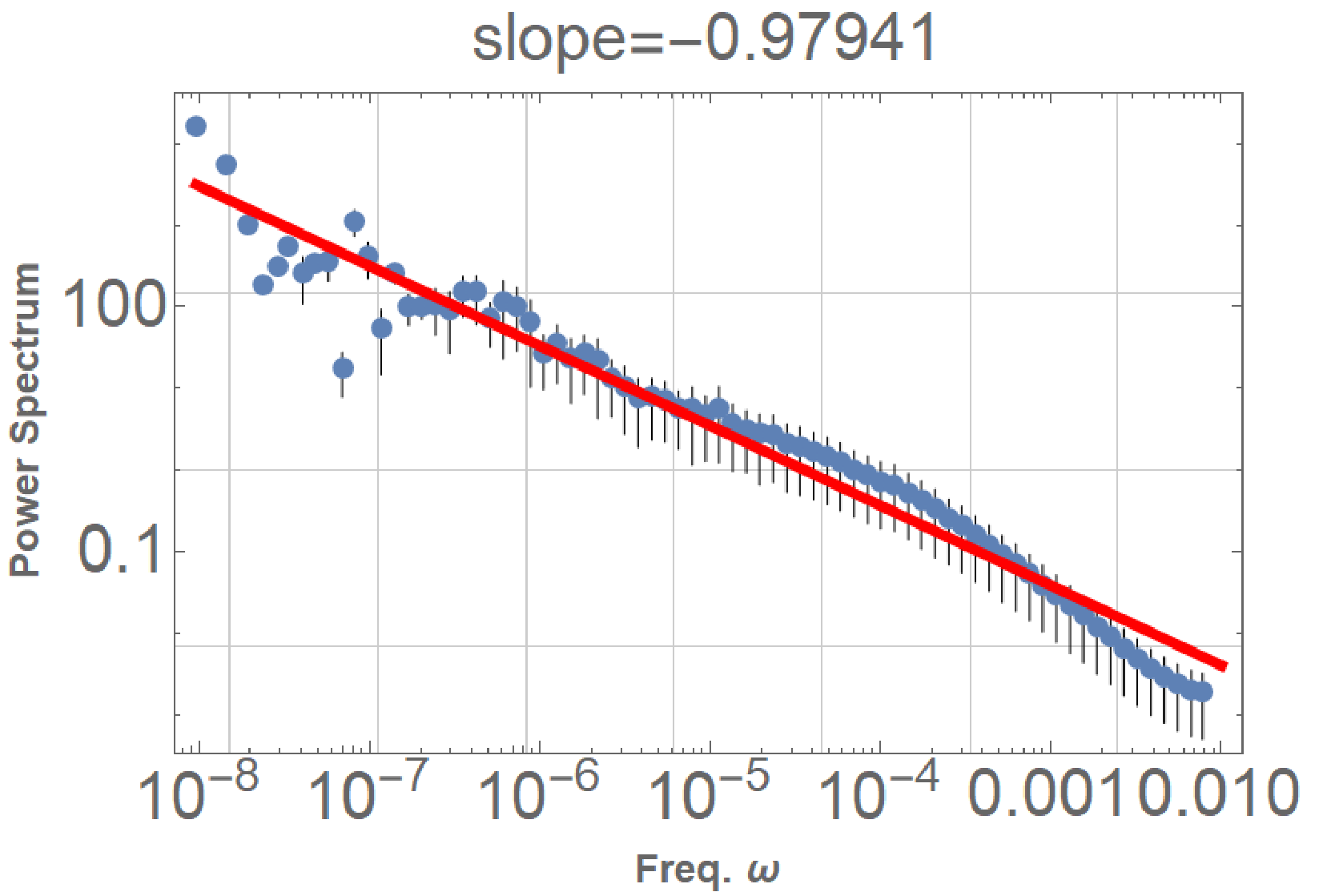}

 \caption{left) Same as Fig.\ref{fig1}, but the energy information is fully
removed; we reset the energy value to one for each data. The PSD shows
the power behavior with an index of -1.1, a typical pink noise, in
the range $10^{-3}$to$2\times10^{-8}$ Hz. \protect \\
right) Same as left PSD but in the high-energy group. This gives an
apparent contrast with the left of Fig.\ref{fig2}, in which all the
energy information is included. }

\label{fig3}
\end{figure}

In order to confirm that the solar flare 1/f fluctuation is independent
from energy, we entirely remove the energy information from the data:
we set all the energy values in the time sequence to one. Then, the
PSD for the entire data turns out to show pink noise with the power
of -1.1 as in Fig\ref{fig3} left. Even the PSD for the high-energy
group, if energy information is likewise removed, shows pink noise
with the power of -0.98 as in Fig\ref{fig3} right.

All together, pink noise with the power index $-0.9\sim\lyxmathsym{\textendash}1.1$
is observed within about five digits of frequencies corresponding
to the timescales from about an hour ($10^{-3}$ Hz) to 6.6 years($2\times10^{-8}$
Hz). This pink noise appears when we remove energetic solar flare
events or fully remove the energy information. 

These facts indicate that the solar flare pink noise does not reflect
the energy scaling structure typically caused by the self-organized
criticality (SOC) formed by energy cascades from small to large, although
SOC may be crucial to explain the popular scaling laws Gutenberg-Richter
and Omori laws. Contrary, the above facts indicate that the solar
flare pink noise is a low-energy phenomenon, probably triggered by
a tiny energy source. 

We have also analyzed short-term GOES16 solar flare data for one week,
arbitrarily chosen. The results are consistent; showing the presence
of pink noise with an index of $-1.15$ within the frequency range
of $2\times10^{-3}$to $2\times10^{-6}$ Hz. This extends across the
entire week, encompassing frequencies lower than the typical frequency
of SFO.

We have further analyzed RHESSI solar flare 16 years data, 2002-2018\citep{RHESSI2018}.
The results are fully consistent, showing the presence of pink noise
with an index of $-1.0$ within the frequency range of $5\times10^{-6}$to
$2\times10^{-9}$ Hz, if we ignore the energy information. This extends
from several days up to the whole observation period.

What mechanism then gives rise to this universal pink noise at the
low-energy regime of various solar flare data?

\section{Amplitude modulation from resonance\label{sec:Amplitude-modulation-and}}

We recently proposed a potential origin for 1/f fluctuation, attributing
it to amplitude modulation \citep{Morikawa2023}. Given the generality
of this mechanism, our intention is to extend its application to the
context of solar flare pink noise within the scope of this paper.

The foundation of this theory rests on the observation that waves
with accumulating frequencies yield robust low-frequency signals.
In cases where this accumulation is systematic, such as in instances
of resonance, synchronization, or infrared divergence, the beats consistently
manifest a power law with an index approximately equal to -1. The
ubiquity of this phenomenon in nature stems from the prevalence of
simple physics governing beat or amplitude modulation.

Consider the example of a beat: waves with frequencies $440$ Hz and
$441$ Hz yield a beat. In musical sounds, this beat is 'audible'
as a sinusoidal amplitude oscillation with a frequency of $1$ Hz.
However, Fourier Transform does not yield a $1$ Hz signal, only the
original two frequencies. To extract this encoded $1$ Hz signal,
a simple method is to square the data and then Fourier Transform.
This allows for the extraction of the encoded low-frequency signal
at 2 Hz, though twice the original. Decoding is not limited to squaring
but can also involve the absolute value, rectification, 4th order
power, thresholding, or other methods, resulting in a variety of pink
noise.

Another example is AM radio, where Amplitude Modulation is utilized.
Using high-frequency radio waves of 526.5 kHz to 1606.5 kHz, a low-frequency
audible signal is encoded. However, the encoded sound cannot be heard
directly, as rapidly oscillating positive and negative parts in the
wave cancel each other out, leaving no audible signal. Demodulation
is achieved by rectifying the radio wave signal, typically through
the use of germanium diodes or vacuum tubes. This process is indispensable
for extracting the encoded low-frequency signal, such as pink noise
\citep{Morikawa2023}.

In the context of pink noise in solar flares, we hypothesize that
the resonant mode crucial for the manifestation of 1/f fluctuations
is the Solar Five-minute Oscillation (SFO), a phenomenon consistently
activated within the solar atmosphere through turbulent convection\citep{Leighton1962,Evans1962}.
Specifically, pressure modes of SFO exhibit accumulating eigenfrequencies,
particularly converging towards lower angular indices $l$. We aim
to examine whether this frequency accumulation effectively produces
a 1/f power spectrum density.

If the SFO induces amplitude modulation, demodulation becomes imperative
for the observation of 1/f fluctuation to occur \citep{Morikawa2023}.
This necessity arises due to the cancellation of positive and negative
components within the relatively high-frequency wave, encompassing
1/f modulation. In the context of solar flares, the demodulation process
is envisioned to be facilitated by the threshold established through
magnetic reconnection. The tiny energy required for magnetic reconnection
may have a 1/f fluctuation characteristic, aligning with the tiny
energy associated with SFO that can trigger solar flares.

Subsequently, we apply this theory of amplitude modulation to the
analysis of 1/f fluctuations in solar flares.

\section{Resonating Solar Five-minute Oscillation \label{sec:Resonator--SFO}}

We delve into the potentiality of the Solar Five-minute Oscillation
(SFO) as a catalyst for 1/f fluctuation in solar flare activity. Specifically,
our focus centers on elucidating how SFO eigenmodes contribute to
the accumulation of frequencies, thereby generating low-frequency
signals through amplitude modulation mechanisms.

The small displacement, denoted as $u(t,r,\theta,\phi)$, of the solar
atmosphere from its equilibrium position follows the Poisson equation

\begin{equation}
\rho\ddot{u}=\kappa\triangle u-\rho\nabla\phi_{g},
\end{equation}
where $p,\rho,\kappa,G,\phi_{g}$ represent the pressure, mass density,
bulk modulus, gravitational constant, and gravitational potential,
respectively. 

The stationary solution $u(t,r,\theta,\phi)=v(r,\theta,\phi)e^{-i\omega t}$
leads to the eigenvalue equation. Utilizing the variable separation
method in the spherical coordinate system, we obtain a solution of
the form
\begin{equation}
u(t,r,\theta,\phi)=R_{n,l,m}(r)Y_{l}^{m}(\theta,\phi)e^{-i\omega_{n,l,m}t},
\end{equation}
where $Y_{l}^{m}(\theta,\phi)$ represents spherical harmonics, and
the modes are characterized by $n=0,1,2,...$, $l=0,1,2...$, and
$-l\leqq m\leqq l$. The modes are further categorized as pressure
and gravitational modes. All parameters are uncertain depending on
the detail of the solar interior, making the solution of the eigenvalue
equation a complex task. Numerous numerical calculations and observational
studies have been conducted on the aforementioned eigenmode equations.

We utilize observational data pertaining to eigenmodes of solar oscillations
obtained through helioseismology \citep{JSOC2023}. This dataset provides
valuable information on numerous observed frequencies, disregarding
the degeneracy in the azimuthal order number parameter $m$ ($-l\leq m\leq l$).
A distinctive characteristic of these modes is the accumulation of
frequencies towards smaller values of $l$ for each $n$ parameter,
typically around $3\times10^{-3}$Hz. This property is pivotal for
the emergence of pink noise through the amplitude modulation mechanism
\citep{Morikawa2023}.

To simulate the phenomenon, we randomly superimpose all sinusoidal
waves with frequencies ranging from the lowest at $848.241\mu$Hz
up to $4669.16\mu$Hz. The wave mode superposition is expressed as
\begin{equation}
\Phi(t)=\sum_{k=1}^{N}\xi_{k}sin(2\ensuremath{\pi}\Omega_{k}t),\label{eq Phi}
\end{equation}
where $\xi_{k}$ is a random variable within the range $[0,1]$, and
$N=2247$ represents the total number of eigenfrequencies in the data.
Subsequently, we conduct Fourier analysis (FFT) on the power spectrum
density (PSD) for the time series of the absolute value $\left|\Phi(t)\right|$.
Notably, calculating PSD for the bare $\Phi(t)$ yields no signal
in the low-frequency domain. As the 1/f fluctuation signal is modulated
in our model, a demodulation process is imperative; taking the absolute
value is a typical demodulation method, essential for extracting 1/f
fluctuations in the PSD analysis. The specifics of the demodulation
process will be explored further in subsequent discussions.

The PSD analysis results in a power-law with an index of approximately$-0.5$
within the low-frequency range of $2\times10^{-5}-5\times10^{-3}$
Hz, as illustrated in Fig. \ref{fig3} left. However, it is noteworthy
that the observed solar flare pink noise occurs in the range of $2\times10^{-8}-10^{-3}$
Hz, considerably lower than our analysis. This disparity suggests
the need for a more nuanced consideration of realistic fine structures
of the eigenstates and additional resonances, a facet we will delve
into in the subsequent analysis.

\begin{figure}
\includegraphics[width=9cm]{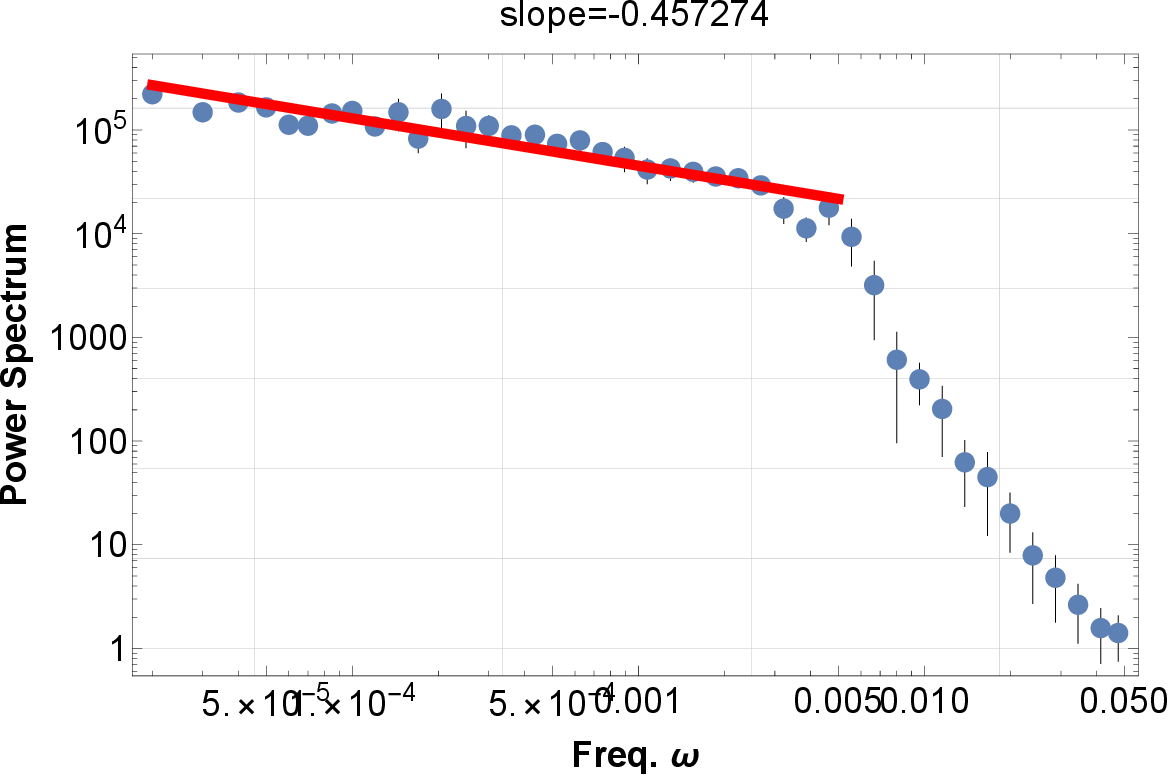}\includegraphics[width=9cm]{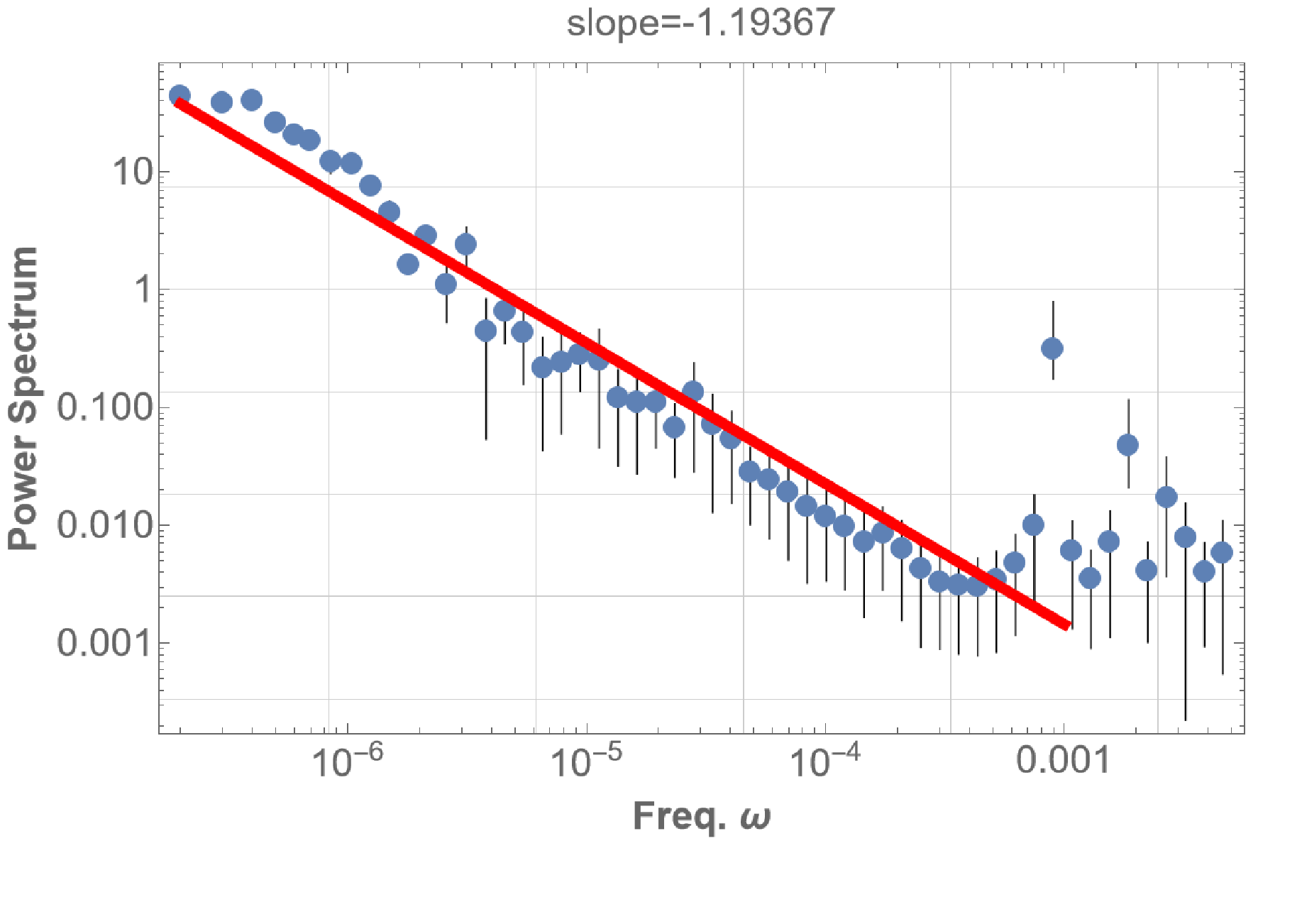}

\textcolor{red}{}

\caption{left: On the left side, the graph depicts the Power Spectrum Density
(PSD) of the absolute value of the time sequence given by Eq.\ref{eq Phi},
denoted as $\left|\Phi(t)\right|$. Here, $\Phi(t)$ represents the
superposition of sinusoidal waves with the $N=2247$ eigenfrequencies
of the Solar Five-minute Oscillation (SFO), each with a random amplitude.
Each mode is identified by the parameters $n,l$, and the azimuthal
order number $m$, which is degenerate. Despite exhibiting a power
law, this presentation barely demonstrates the characteristics of
pink noise. \protect \\
right: On the right side, analogous to the left graph, this graph
includes resonant modes and fine eigenmodes after resolving the degeneracy
in $m$. In constructing the data $\Phi(t)$, we superimpose sinusoidal
waves with 100 frequencies from the lowest and introduce $N=100$
Lorentzian-distributed modes. The latter are randomly generated following
Eq.\ref{eq:R}. The graph displays the PSD of the thresholded value
of the time sequence Eq.\ref{eq:Phi2},$\left|\Phi(t)\right|$. The
threshold is strategically set to select data points $\left|\Phi(t)\right|$
that surpass twice the mean, although insensitive to the demodulation
method. This presentation reveals nearly 1/f fluctuation with an index
of $-1.2$ spanning over four digits. Notably, variations in thresholds
and sample sizes in the PSD analysis consistently yield similar 1/f
fluctuations.}
\label{fig4}
\end{figure}

Our analysis of resonance is currently incomplete, with several aspects
requiring further exploration. Firstly, a) each eigenmode is associated
with a resonance curve, and numerous frequency-accumulating modes
are linked to each mode. Secondly, b) the degeneracy in the azimuthal
order number $m$ should give rise to a fine structure around each
principal frequency characterized by $n$ and $l$. This degeneracy
in $m$ is resolved by the solar non-spherical symmetry or the solar
rotation. In this paper, we examine representative modes for both
cases (a and b) to illustrate how the fine structure contributes to
the emergence of 1/f fluctuation. A comprehensive analysis encompassing
these aspects will be presented in our forthcoming publications.

To refine the Power Spectrum Density (PSD), we introduce the following
effects: a) each eigenfrequency labeled by $n$ and $l$ possesses
a finite width, and b) the degeneracy in $m$ is resolved by the solar
rotation.

a) Resonant modes are typically modeled by the Lorentzian distribution,
\begin{equation}
R[\omega]=\frac{1}{\left(\frac{\kappa}{2}\right)^{2}+\left(\omega-\Omega\right){}^{2}},\label{eq:R}
\end{equation}
where $\Omega$ is the fiducial resonance frequency, and $\kappa$
characterizes the sharpness of the resonance. This function represents
the frequency distribution density associated with the fiducial frequency
$\Omega$. The inverse function (tangent) of the cumulative distribution
function (hyperbolic tangent) generates this distribution from the
Poisson random field.

b) Solar rotation resolves the degeneracy in $m$ by breaking the
spherical symmetry of the system. Although the details are intricate,
a rough estimate is provided by the resolved frequency \citep{Duvall1984}
in the lowest perturbation in $\Omega/\omega$ $(\ll1)$, 
\begin{equation}
\omega_{nlm}=\omega_{nl}+\frac{m}{l(l+1)}\Omega,\label{eq:Omega}
\end{equation}
where $\omega_{nl}$ is the degenerate eigenfrequency, and $\Omega=4.3\times10^{-7}Hz$
is the frequency associated with solar rotation. The coefficient of
$\Omega$ is chosen approximately according to \citep{Duvall1984}.

These effects are implemented through a specific process. Initially,
we construct wave data by superposing $N$ sinusoidal waves with eigenfrequencies
after eliminating the degeneracy in $m$. Additionally, we superimpose
$M$ resonant waves with frequencies proximate to the fiducial frequency,
following the distribution in Eq.(\ref{eq:R}). The fully superposed
wave is defined as:

\begin{equation}
\Phi(t)=\sum_{n=1}^{N}\sum_{i=1}^{M}\sin\left(2\pi(1+c\tan(\xi_{i}))\Omega_{n}t\right),\label{eq:Phi2}
\end{equation}
where the parameter $c=\kappa/\Omega$ represents the relative line
width for each eigenfrequency. The random variable $\xi_{i}$, ranging
in $[0,\pi/2]$, generates the frequency distribution through $R(\omega)$
in Eq.(\ref{eq:R}). While $c$ actually depends on each $n$, for
simplicity, we use $c=0.01$. We utilized data \citep{JSOC2023},
limiting $M$ to $100$ and $N$ to $100$.

As before, the power spectrum density (PSD) of the bare $\Phi(t)$
exhibits no signal in the low-frequency region. However, taking the
absolute value $\left|\Phi(t)\right|$ or applying arbitrarily set
threshold data produces 1/f fluctuations (details in the caption of
Fig.\ref{fig4}). These square operations and thresholding essentially
function as a demodulation of the original signal. Consequently, the
1/f fluctuation becomes evident only after demodulation and proves
to be quite robust. Figure \ref{fig4} right illustrates the PSD of
the thresholded data, demonstrating an approximate 1/f fluctuation
with a power index of$-1.2$, covering a frequency range extended
down to $2\times10^{-7}$Hz. This range partially coincides with the
observed range below $10^{-3}$ Hz. In our future study, further refinement
of the PSD analysis is intended, incorporating finer structures of
eigenfrequencies, decay times, and deviations of the Sun from spherical
symmetry. The introduction of gravitational modes, alongside pressure
modes, which operate in much lower frequency domains, is also of interest.

In the preceding discussion, we superimposed the eigenmodes of the
Solar Five-minute Oscillation (SFO) to obtain amplitude modulation
and pink noise, explaining the observed pink noise in solar flares.
However, a more direct examination of the bare data of SFO before
decomposition into eigenmodes is warranted. This implies that the
resonance of SFO directly yields pink noise. Initially, we utilize
SOHO-GOLF data on the fluctuations of the time elapsed $T(t)$ for
the waves to circumnavigate the Sun \citep{Boumier2023}. Details
are expounded in \citep{Fossat2017}, with the data spanning about
$16.5$ years and an interval of $80$ seconds, albeit with some data-missing
periods.

\begin{figure}
\includegraphics[width=9cm]{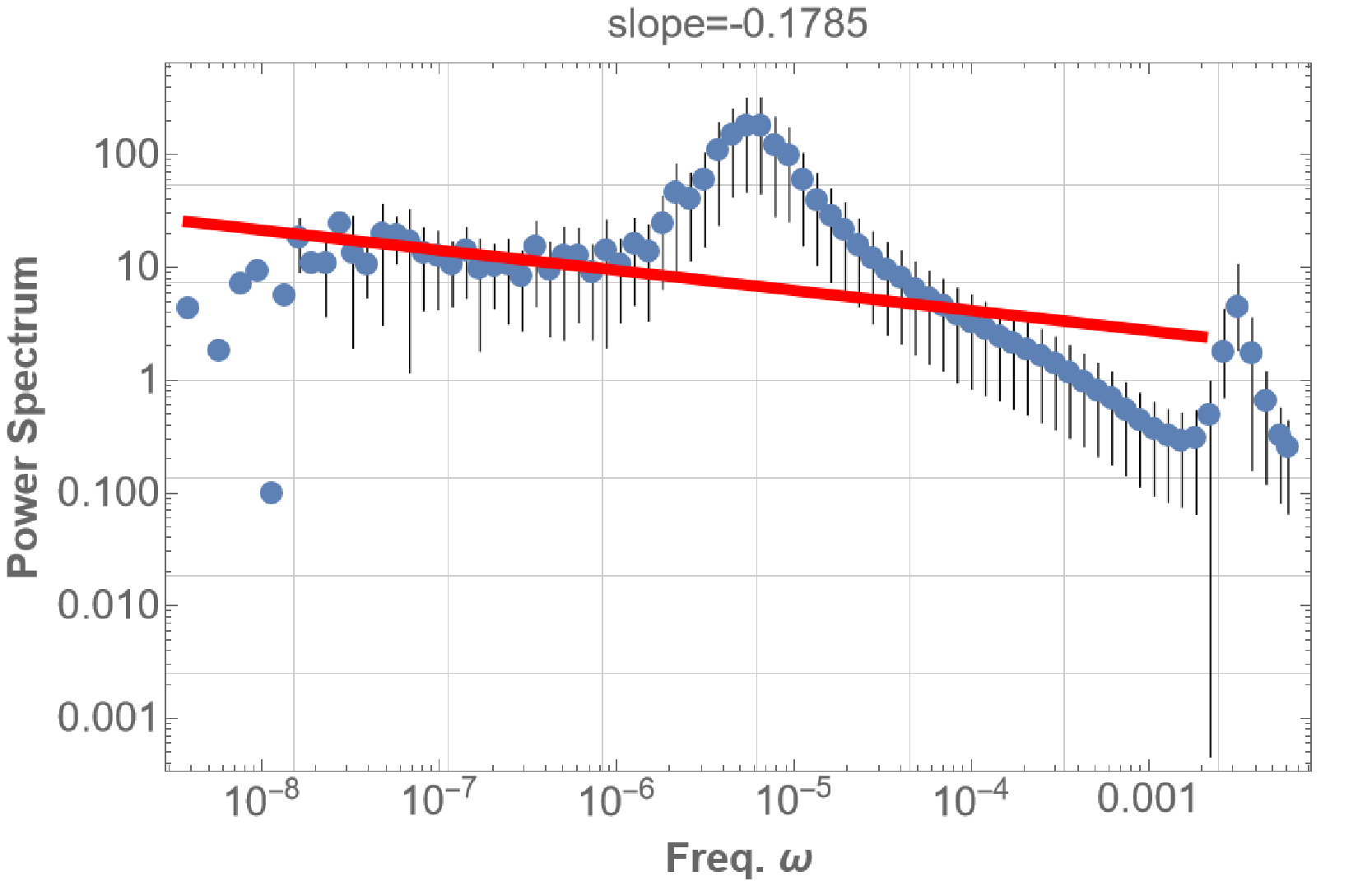}\includegraphics[width=9cm]{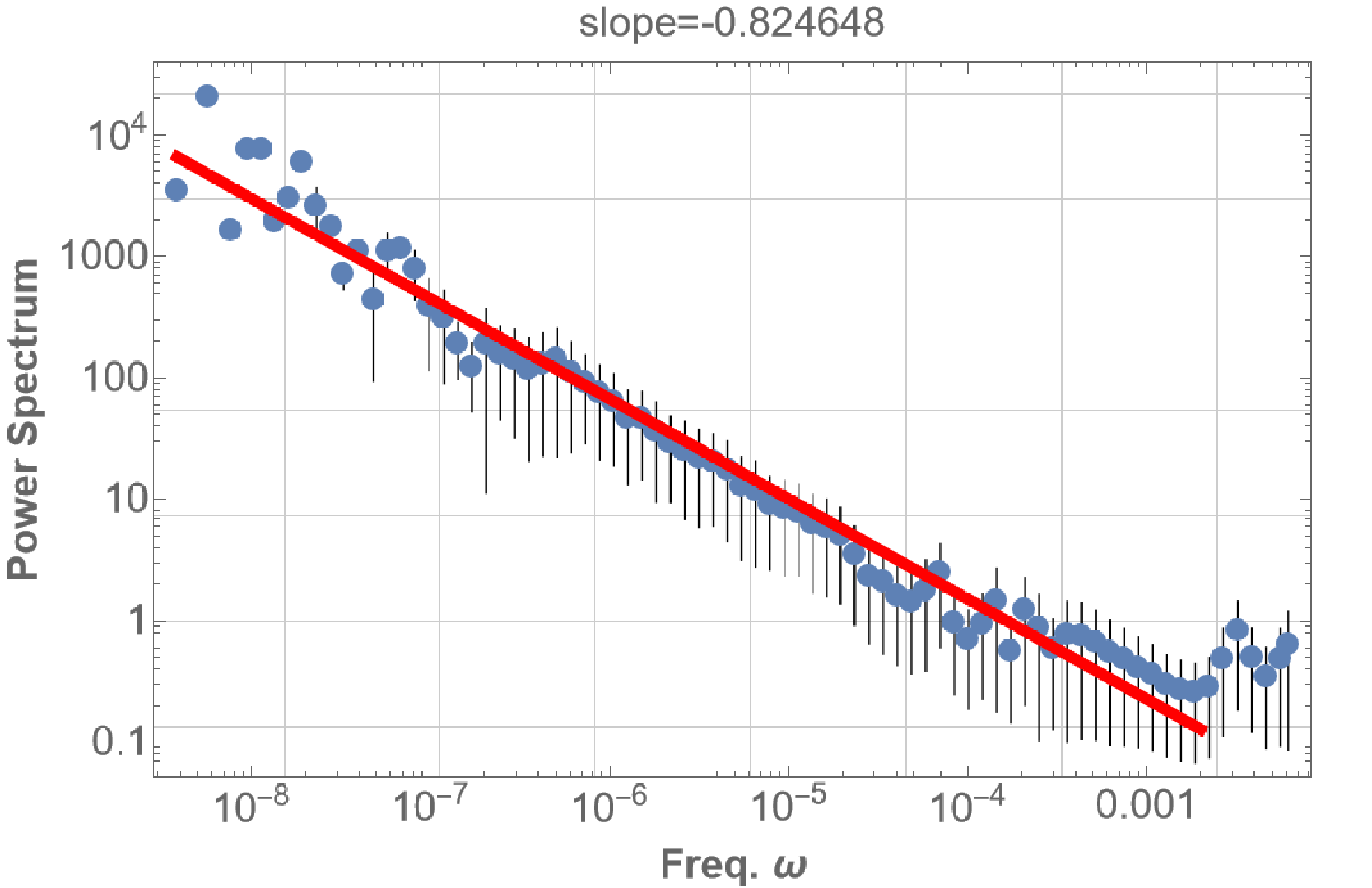}\caption{left) The Power Spectrum Density (PSD) is presented for the SOHO-GOLF
16.5-year data, focusing on the time elapsed T(t) for the waves circumnavigating
the sun \citep{Boumier2023}. The rightmost peak corresponds to the
typical five-minute mode of the Solar Five-minute Oscillation (SFO)
at $3\times10^{-3}$ Hz. \protect \\
right) The PSD is shown for the absolute value of the original data,
revealing pink noise with a slope of $-0.8$ over approximately six
digits.}
\label{fig5}
\end{figure}

We commence by calculating the Power Spectrum Density (PSD) of the
original data $T(t)$. The result is depicted in the left graph of
Fig.\ref{fig5}. A prominent peak appears around $3\times10^{-3}$Hz,
corresponding to the typical five-minute mode of solar oscillation.
From there, a partial power-law behavior is observed toward $6\times10^{-6}$Hz
with an index of about$-1$, followed by a flat behavior at lower
frequencies. This partial pink noise may stem from instrumental origins
\citep{Gabriel1995}, potentially not reflecting genuine solar properties.

Conversely, when taking the absolute value of the data, the pink noise
region in PSD extends toward the lowest frequency limit, as illustrated
in Fig.\ref{fig5} right. The behavior of the SOHO-GOLF data $T(t)$
is precisely the same as the sound data of orchestra music \citep{Morikawa2021}
or the sound of a big bell. While the sound wave amplitude time-sequence
data itself does not exhibit pink noise, the square of the amplitude
showcases apparent pink noise. These operations of squaring or taking
the absolute value naturally correspond to the demodulation process,
revealing the encoded pink noise.

This forms part of the rationale behind our belief that solar eigen-oscillations
contribute to the pink noise observed in solar flares.

An alternative analysis involves ground-based data from BiSON \citep{Davies2014,Hale2016},
measuring radial velocity from January 1985 to January 2023 (All sites,
Optimized for Quality). The PSD of the original bare data does not
exhibit pink noise. However, taking the absolute value of the original
data results in clear pink noise with a slope of $-0.71$, although
this index is slightly larger than the right side of Fig.\ref{fig5}.
This discrepancy is likely influenced by artificial peaks corresponding
to the periods of the Earth's spin and rotation at $1.1\times10^{-5}$
Hz and $3\times10^{-8}$Hz.

A variety of demodulation methods has been considered in association
with this BiSON data manipulation. In our previous analysis, we focused
on the absolute value of the data, but it is noteworthy that other
manipulations can also be employed to extract pink noise. For instance,
the squared data exhibits pink noise, while the pink noise disappears
in the cubed data. However, when the data is raised to the fourth
power, pink noise reappears. These findings strongly indicate that
the amplitude-modulated pink noise emerges after specific demodulation
processes.

Similar phenomena are often observed in sound systems. For example,
we examined sound data collected at the water-harp cave (Suikinkutsu)
at HosenIn Temple in Kyoto \citep{HosenIn2022}. The sound is generated
by the perpetual impact of water drops on the water surface in the
two-meter Mino-yaki pot underground \citep{Suikinkutsu2023}. Although
the original sound data barely shows pink noise, the squared data
from this sound source clearly displays 1/f fluctuation with an index
of -0.80 for four digits. The resonator, in this case, is presumed
to be the Mino-yaki pot.

\section{Timing Statistics\label{sec:V. Timing Statistics}}

In our analysis of Sec.\ref{II. Solar flare fluctuations}, we identified
pink noise in the time series of solar flare timing and speculated
that this characteristic might be indicative of the low-energy trigger
for solar flares. In seismic activity, which also exhibits pink noise\citep{Nakamichi2023},
the time series of earthquake occurrences is often described by the
Weibull distribution function \citep{Tanaka2017,Hatano2017}:

\begin{equation}
f(x)=\frac{\alpha}{\beta}\left(\frac{x}{\beta}\right)^{\alpha-1}\exp\left(-\left(\frac{x}{\beta}\right)^{\alpha}\right).
\end{equation}
We now explore the extent to which the solar flare timing time sequence
is characterized by the Weibull distribution and its potential connection
to pink noise.

\begin{figure}
\includegraphics[width=9cm]{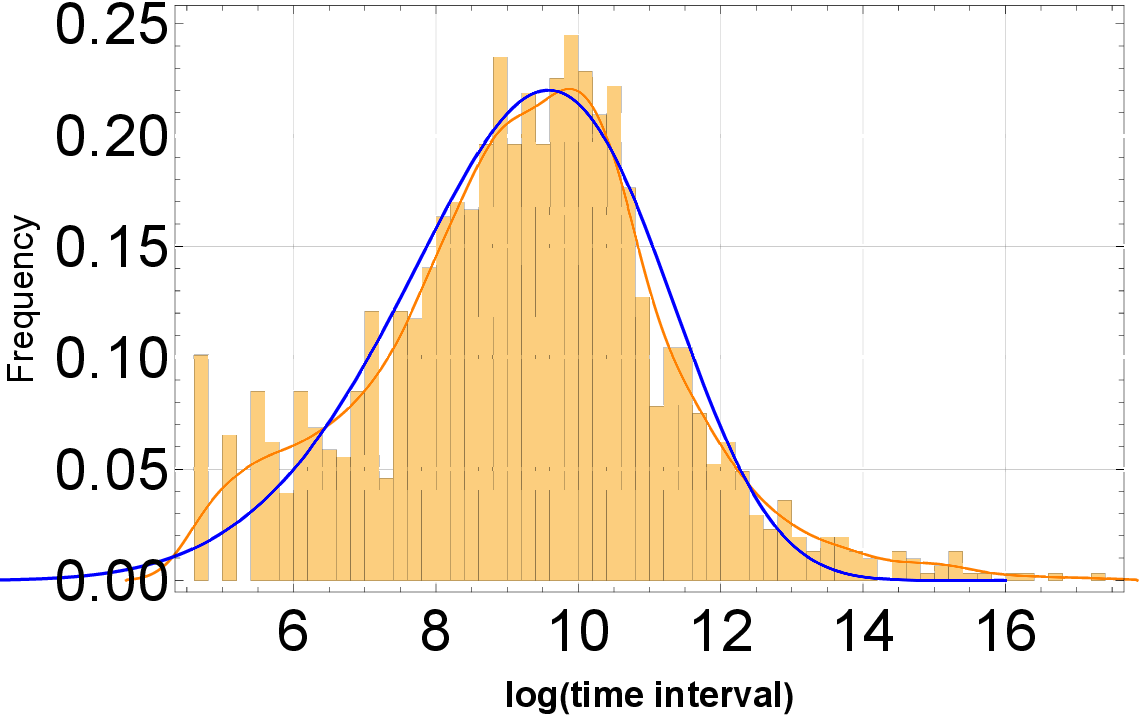}\includegraphics[width=9cm]{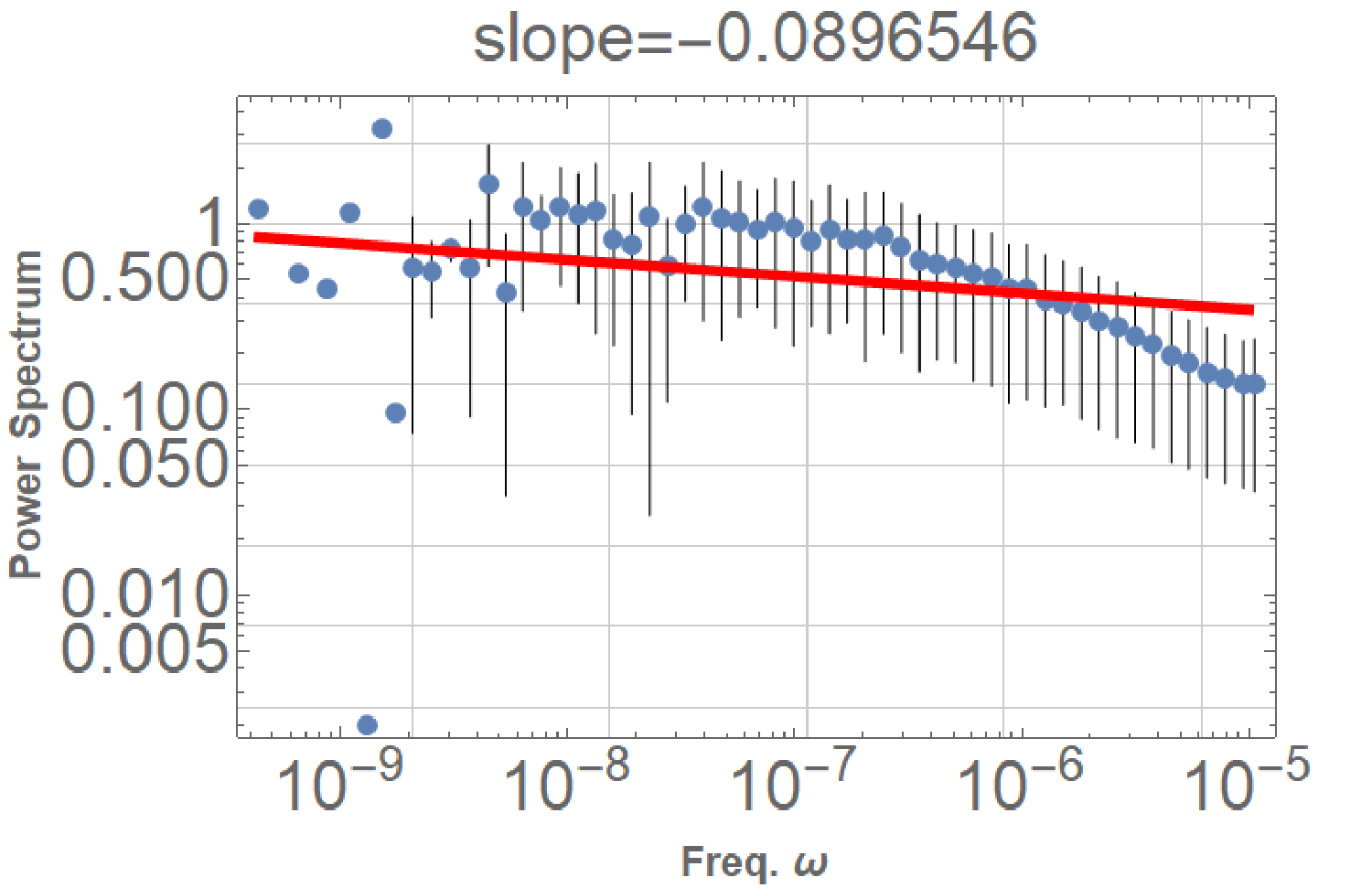}

\caption{left) Orange bars represent the frequency distribution of the logarithm
of the time intervals between solar flare occurrences. The blue line
denotes the Weibull distribution with the parameters $\alpha=5.3$
and $\beta=9.9$, best fitting the solar flare data. \protect \\
right) The PSD is presented for artificial data with random time intervals
generated by the best-fit Weibull distribution. It is evident that
the PSD of this artificial data is flat, indicative of a random distribution. }
\label{fig6}
\end{figure}

Upon examination of the GOES data \citep{GOES2023} used in our analysis,
we discover that, in the logarithm of the time interval, it follows
the Weibull distribution, as illustrated in Fig. \ref{fig6} left.
The best-fit parameters are $\alpha=5.3$ and $\beta=9.9$. Therefore,
similar to seismic activity, the statistical distribution of solar
flare timing can be effectively characterized by the Weibull distribution.
The question then arises: to what extent does this Weibull distribution
characterize the pink noise property?

We have checked that a time sequence simply following the Weibull
distribution does not exhibit pink noise; the Power Spectrum Density
(PSD) becomes flat in the low-frequency range, as depicted in Fig.
\ref{fig6} right. Consequently, the pink noise observed in solar
flare timing is independent of the Weibull distribution, as the behavior
observed in seismic cases. This property is readily understood; the
time sequence, constructed with randomly chosen intervals according
to Weibull distribution statistics, lacks long correlation times.
In contrast, pink noise inherently possesses long correlation times,
thereby manifesting as a low-frequency property.

\section{Robustness and Inherited Pink Noise \label{sec:VI. Robustness and Inherite Pink Noise}}

We have delved into the origin of pink noise in solar flares, applying
the overarching concept that the accumulation of frequencies in numerous
waves leads to beat or amplitude modulation. Frequency accumulation
is achieved through resonance, and we have successfully explored this
concept using the Solar Five-minute Oscillations (SFO), which are
consistently resonating. It is plausible that magnetic reconnection
serves as the demodulation (DM) of this amplitude modulation (AM),
resulting in pink noise in solar flares. If this holds true, the combination
of SFO (as AM) and magnetic reconnection (as DM) may generate pink
noise beyond solar flare in other extended regions of the solar neighborhood.

For instance, if magnetic reconnection induces solar wind through
jetlets \citep{Raouafi2023}, triggered by SFO, then the solar wind
\citep{Parker1958} may also exhibit pink noise. Indeed, pink noise
in the solar wind has been observed over many years\citep{Verdini2012,Matteini2019,Huang2023}.
SFO has been detected in the solar corona \citep{Didkovsky2012},
suggesting the potential observation of pink noise in that context
as well.

Furthermore, the solar wind may interact with the Earth's atmosphere,
inducing chemical reactions leading to the production of $\mathrm{NO_{3}^{-}}$
isotope. This isotope could then become embedded in Antarctic ice
cubes. This work is in progress \citep{Motizuki2023}. If solar wind
and solar flares influence the Earth's surface, then the sea surface
temperature may also exhibit pink noise \citep{Koyama2023}, similar
to the case of seismic activity.

In seismic activity, which also displays pink noise, seismic events
inherit a pink noise pattern, potentially resulting from amplitude
modulation due to resonance with Earth Free Oscillation (EFO) in the
lithosphere \citep{Nakamichi2023}. This EFO may further contribute
to the pink noise observed in the time sequences of volcano eruptions
and the fluctuation of the Earth's rotation axes \citep{Nakamichi2023-1}.

Including the above robustness of pink noise, we compare solar flares
and earthquakes in Table \ref{table1}. This table is preliminary
and will be finalized in our forthcoming study.

\begin{table}

\begin{tabular}{|c|c|c|}
\hline 
 & Solar Flare & Eathquakes\tabularnewline
\hline 
\hline 
AM (resonator) & Solar Five-minute Oscillation (SFO) & Earth Free Oscillation (EFO)\tabularnewline
\hline 
DM & magnetic reconnection & fault rupture\tabularnewline
\hline 
PSD total data & flat (GOES16, RHESSI) & flat (USGS)\tabularnewline
\hline 
PSD low-energy & pink (GOES16, RHESSI) & pink (USGS)\tabularnewline
\hline 
PSD timing & pink (GOES16, RHESSI) & pink (USGS)\tabularnewline
\hline 
PSD superposed eigenmodes & pink (JSOC) & pink (T. G. Masters, R. Widmer )\tabularnewline
\hline 
Weibull Distribution & yes: $\alpha=5.3$ and $\beta=9.9$ (GOES16) & yes: $\alpha=6.3$ and $\beta=7.63$ (USGS)\tabularnewline
\hline 
PSD resonator & pink (SOHO-GOLF, BiSON) & ?\tabularnewline
\hline 
inherent phenomena & \multirow{1}{*}{solar wind, sunspot number, nitrate, SST,cosmic ray\citep{Koyama2023}} & volcano eruption, rotation axes\citep{Nakamichi2023-1}\tabularnewline
\hline 
\end{tabular}

\caption{Similarity of solar flares and earthquakes from the view point of
pink noise. \protect \\
This is a tentative table, and the detail will be reported soon by
the authors. }

\label{table1}
\end{table}

\section{Conclusions and prospects \label{sec:VII. Conclusions and prospects}}

In conclusion, our investigation has identified pink noise in the
solar flare time series, and we have partially elucidated its origin
by applying our proposed mechanism: pink noise emerges from the resonance
of the Solar Five-minute Oscillations via amplitude modulation and
demodulation.

Our GOES data analysis of the solar flare time sequence has revealed
distinct low-frequency properties. Specifically, the power spectrum
density of low-energy flares ($E\le E_{mean}$) exhibited 1/f fluctuations,
while high-energy flares ($E>E_{mean}$) displayed a flat spectrum.
Notably, the time sequence of flare occurrences demonstrated clearer
1/f fluctuations, indicating that low-energy characteristics play
a pivotal role in triggering the observed 1/f fluctuations in solar
flares. Building on our recent proposal that 1/f noise arises from
amplitude modulation and demodulation, we postulated that this modulation
is encoded through resonance with the Solar Five-minute Oscillation
(SFO) and demodulated via magnetic reconnection.

To test this hypothesis, we constructed a dataset by superposing sinusoidal
waves with $2247$ eigenfrequencies of SFO. The absolute value of
this time sequence marginally exhibited 1/f fluctuations with a power
index of $-0.5$ down to $2\times10^{-5}$Hz. Further refinement of
the data, considering resonance effects and finer structures labeled
by $m$ induced by solar rotation, involved adding $100$ extra modes
generated by resonant Lorentzian distributions for the first $100$
eigenfrequencies of SFO after resolving the degeneracy in m. The absolute
value of this refined time sequence clearly displayed 1/f fluctuations
with a power index of $-1.2$ down to $10^{-7}$Hz, largely overlapping
with the observed range of solar flare 1/f fluctuations (power index
$-1.0$ from about $2\times10^{-3}$ Hz down to$2\times10^{-8}$Hz).
Thus, our analysis provided partial verification that SFO triggers
seismic 1/f fluctuations.

Further investigation into SOHO-GOLF data and BiSON data for velocity
fluctuations in the solar atmosphere revealed that the original time
sequence of these data barely exhibited 1/f fluctuations, while the
absolute values of the time sequence did display clear 1/f fluctuations.
This lends additional support to our proposition: SFO is the origin
of solar flare 1/f fluctuations.

Additionally, our examination of the time sequence of solar flare
occurrences revealed adherence to a Weibull distribution. However,
an artificial time sequence composed from the Weibull distribution
barely exhibited 1/f fluctuations, suggesting that the Weibull distribution
does not fully characterize solar 1/f fluctuations.

Lastly, a comparison between 1/f fluctuations in solar flares and
earthquakes demonstrated remarkable similarities \citep{Kossobokov2008}\citep{Arcangelis2008}.
Furthermore, we propose that a comparable analysis may be applicable
to the activity of a black hole/disk system, replacing the resonator
SFO with the Quasi-Periodic Oscillation of a black hole. This broader
perspective underscores the potential universality of the proposed
amplitude modulation and demodulation mechanism in diverse astrophysical
phenomena.
\begin{acknowledgments}
We extend our heartfelt gratitude to the members of the Lunch-Time
Remote Meeting and the Department of Physics at Ochanomizu University,
as well as Ms. Izumi Uesaka and Mr. Manaya Matsui at Kyoto Sangyo
University, for their engaging discussions and unwavering support.
Special appreciation goes to Professor Tim Larson and Professor Takashi
Sekii for their invaluable insights into solar data. Additionally,
we express our sincere thanks to Professor Satoru Ueno for enlightening
discussions on the topic of solar flare pink noise. Their contributions
have enriched our work and made this research journey a truly collaborative
and meaningful experience. 
\end{acknowledgments}


\begin{thebibliography}{99}
\bibitem{Benz2008}Arnold O. Benz, \textit{Flare Observations}, Living
Rev. Solar Phys., 5, (2008), 1., http://www.livingreviews.org/lrsp-2008-1

\bibitem{Gutenberg1944}B. Gutenberg and C. F. Richter, \textit{Frequency
of Earthquakes in California}, Bull. Seismol. Soc. Am. 34, 185, 1944.

\bibitem{Omori1894}F. Omori, \textit{On the Aftershocks of Earthquakes},
J. Coll. Sci., Imp. Univ. Tokyo 7, 111 (1894).

\bibitem{Arcangelis2006} L. de Arcangelis, C. Godano, E. Lippiello,
and M. Nicodemi, \textit{Universality in Solar Flare and Earthquake
Occurrence}, Phys. Rev. Lett. 96, 051102, 2006.

\bibitem{Najafi2020} Amin Najafi ,\textit{The Modified Form of the
Gutenberg-Richter Law in Solar Flare Complex Network : Approach of
Genetic Algorithm on the Thresholded Power-Law Behavior}, Iranian
Journal of Astronomy and Astrophysics 7-1, 2020. \\
 

\bibitem{Raychaudhuri2002} A.K. Raychaudhuri, \textit{Measurement
of 1/f noise and its application in materials science}, Current Opinion
in Solid State and Materials Science 6-1, 67-85, 2002. https://doi.org/10.1016/S1359-0286(02)00025-6

\bibitem{Milott2002}Edoardo Milott, \textit{1/f noise: a pedagogical
review}. https://arxiv.org/ftp/physics/papers/0204/0204033.pdf \\
arXiv:physics/0204033. 

\bibitem{Morikawa2023}M. Morikawa, A. Nakamichi,\textit{ A simple
model for pink noise from amplitude modulations}, Scientific Reports
13, 8364, (arXiv:2301.11176), 2023. 

\bibitem{Nakamichi2023}Nakamichi, Akika, Matsui, Manaya, Morikawa,
Masahiro, \textit{Seismic 1/f Fluctuations from Amplitude Modulated
Earth's Free Oscillation, }arXiv:2307.03192.

\bibitem{Leighton1962}Leighton, R. B.; Noyes, R. W.; Simon, G. W.,
\textit{Velocity Fields in the Solar Atmosphere. I. Preliminary Report},
Astrophysical Journal, 135: 474. 1962. 

\bibitem{Evans1962}Evans, J. W.; Michard, R., \textit{Observational
Study of Macroscopic Inhomogeneities in the Solar Atmosphere. III.
Vertical Oscillatory Motions in the Solar Photosphere}, Astrophysical
Journal, 136: 493, 1962. 

\bibitem{GOES2023}NOAA, National Centers for Environmental Information,
\\
https://data.ngdc.noaa.gov/platforms/solar-space-observing-\textbackslash{}
satellites/goes/goes16/l2/data/xrsf-l2-avg1m\_science/\\

\bibitem{Ueno1997}S.Ueno, S.Mineshige, H.Negoro, K.Shibata, and
H.S.Hudson \textit{Statistics of Fluctuations in the Solar Soft X-ray
Emission}, Astrophys. J.,, 484, 920-926, 1997. 

\bibitem{RHESSI2018}The Reuven Ramaty High Energy Solar Spectroscopic
Imager (RHESSI), https://hesperia.gsfc.nasa.gov/rhessi3/\\

\bibitem{JSOC2023}Joint Science Operations Center (JSOC), \\
http://jsoc.stanford.edu/\\

\bibitem{Duvall1984} T. L. Duvall Jr, J. W. Harvey, \textit{Rotational
frequency splitting of solar oscillations}, Nature 310 5 1984. 

\bibitem{Gilbert1965}F. Gilbert and G. Backus, \textit{The Rotational
Splitting of the Free Oscillations of the Earth, 2}, Reviews of geophysics
3, 1-9, 1965. 

\bibitem{Boumier2023}GOLF Patrick Boumier \\
https://www.ias.u-psud.fr/golf/templates/access.html\\

\bibitem{Fossat2017} E. Fossat, P. Boumier, T. Corbard, J. Provost,
D. Salabert, et al.. \textit{Asymptotic g modes: Evidence for a rapid
rotation of the solar core}. Astronomy and Astrophysics, 2017, 604,
pp.A40.  \\

\bibitem{Gabriel1995}Gabriel, A. H. ; Grec, G. ; Charra, J. ; Robillot,
J. -M. ; Roca Cortés, T. search by orcid ; Turck-Chièze, S. ; Bocchia,
R. ; Boumier, P. ; Cantin, M. ; Cespédes, E. ; Cougrand, B. ; Crétolle,
J. ; Damé, L. ; Decaudin, M. ; Delache, P. ; Denis, N. ; Duc, R. ;
Dzitko, H. ; Fossat, E. ; Fourmond, J. -J.\textit{, Global Oscillations
at Low Frequency from the SOHO Mission (GOLF)} ; ... Solar Physics,
162, 1-2, pp. 61-99 1995  \textit{}

\bibitem{Morikawa2021}Masahiro Morikawa, \textit{Low-Frequency Characterization
of Music Sounds -{}- Ultra-Bass Richness from the Sound Wave Beats},
arxiv arXiv:2104.08872. https://arxiv.org/abs/2104.08872

\bibitem{Davies2014} \textit{G. R. Davies,\guilsinglleft{} W. J.
Chaplin, Y. Elsworth and S. J. Hale, BiSON data preparation: a correction
for differential extinction and the weighted averaging of contemporaneous
data}, Monthly Notices of the Royal Astronomical Society, 441-4, 2014,
pp3009\textendash 3017.  

\bibitem{Hale2016}S. J. Hale, R. Howe, W. J. Chaplin, G. R. Davies
\& Y. P. Elsworth ,\textit{ Performance of the Birmingham Solar-Oscillations
Network (BiSON) }291, pp1\textendash 28 (2016).

\bibitem{HosenIn2022}HosenIn at Kyoto, http://www.hosenin.net/

\bibitem{Suikinkutsu2023}Suikinkutsu http://www.suikinkutsu.com/ 

\bibitem{Tanaka2017}Hiroki Tanaka and Yoji Aizawa, \textit{Detailed
Analysis of the Interoccurrence Time Statistics in Seismic Activity},
J. Phys. Soc. Jpn. 86, 024004, 2017. 

\bibitem{Hatano2017}Takahiro Hatano, \textit{The Third Law of Earthquake
Statistics?}, JPSJ News Comments 14, 03, 2017. 

\bibitem{Kossobokov2008}Vladimir G. Kossobokov, Fabio Lepreti, and
Vincenzo Carbone,\textit{ Complexity in Sequences of Solar Flares
and Earthquakes,} Pure appl. geophys. 165 (2008) \\

\bibitem{Arcangelis2008}L. de Arcangelis1, E. Lippiello, C. Godano,
and M. Nicodemi, \textit{Statistical properties and universality in
earthquake and solar flare occurrence,} Eur. Phys. J. B 64, 551\textendash 555
(2008) \\
 

\bibitem{Didkovsky2012} L. Didkovsky1, A. Kosovichev, D. Judge1,
S. Wieman, T. Woods\textit{, Variability of Solar Five-Minute Oscillations
in the Corona as Observed by the Extreme Ultraviolet Spectrophotometer
(ESP) on the Solar Dynamics Observatory Extreme Ultraviolet Variability
Experiment (SDO/EVE) }arXiv:1211.0711 2012.\textit{ }

\bibitem{Raouafi2023}\textit{}Nour E. Raouafi1, G. Stenborg, D.
B. Seaton, H. Wang, J. Wang, C. E. DeForest, S. D. Bale, J. F. Drake,
V. M. Uritsky, J. T. Karpen\textit{, et al. Magnetic Reconnection
as the Driver of the Solar Wind, }The Astrophysical Journal, 945-1
28, 2023. 

\bibitem{Parker1958}Eugene Newman Parker\textit{, Dynamics of the
Interplanetary Gas and Magnetic Fields}, Astrophysical Journal 128:664,
1958.

\bibitem{Matteini2019}Matteini, L. ,Chen, C. H. K. ,Stansby, D.
,Horbury, et al.  \textit{Large scale 1/f magnetic field spectrum
in the solar wind close to the Sun: comparison between 0.15 and 0.3AU}
\\
American Geophysical Union, Fall Meeting 2019, abstract SH21C-3329
2019. 

\bibitem{Verdini2012}\textit{}A. Verdini, R. Grappin, R. Pinto,
M. Velli, \textit{On the origin of the 1/f spectrum in the solar wind
magnetic field }Solar and Stellar Astrophysics, arXiv:1203.6219  

\bibitem{Huang2023}\textit{}Zesen Huang et.al. \textit{, New Observations
of Solar Wind 1/f Turbulence Spectrum from Parker Solar Probe,} https://arxiv.org/pdf/2303.00843.pdf

\bibitem{Motizuki2023}Y. Motizuki and M. Morikawa,\textit{ pink noise
in NO3-,} in preparation. 

\bibitem{Koyama2023}T. Koyama, Y. Motizuki and M. Morikawa, \textit{pink
noise in SST,} in preparation. 

\bibitem{Nakamichi2023-1}A. Nakamichi, and M. Morikawa, pink noise
around seismic activities, in preparation. 
\end{thebibliography}
\end{document}